\documentclass[twocolumn,aps,epsfig]{revtex4}
\usepackage{epsfig}
\newcommand{\gsim}{\mathrel{\raisebox{-.6ex}{$\stackrel{\textstyle>}{\sim}$}}}
\begin{document}

\markboth{Dutta, Reno, Sarcevic}
{Ultrahigh Energy Neutrinos}

\title{ULTRAHIGH ENERGY NEUTRINOS}

\author{ SHARADA IYER DUTTA}

\address{Department of Physics and Astronomy, State University of New York
at Stony Brook\\
Stony Brook, New York 11794, USA
}

\author{MARY HALL RENO}\altaffiliation[Presented at ]{{\it Neutrinos and 
Implications for Physics Beyond the Standard Model}, Stony Brook, NY, October 11-13, 2002.}

\address{Department of Physics and Astronomy, University of Iowa\\
Iowa City, Iowa 52242, USA
}

\author{INA SARCEVIC}

\address{Department of Physics, University of Arizona\\
Tucson, Arizona 85721, USA}

%\pub{Received (14 January 2003)}{ }

\begin{abstract}
The ultrahigh energy neutrino cross section is well understood in
the standard model for neutrino energies up to $10^{12}$ GeV.
Tests of neutrino oscillations $(\nu_\mu\leftrightarrow \nu_\tau)$
from extragalactic sources of neutrinos are possible with large
underground detectors. Measurements of horizontal air shower event
rates at neutrino energies above $10^{10}$ GeV will be able to constrain
nonstandard model contributions to the neutrino-nucleon cross section,
{\it e.g.}, from mini-black hole production.

\end{abstract}

\maketitle

\section{Introduction}	%) A SECTION HEADING

The existence of ultrahigh energy (UHE) cosmic rays comprised of
nucleons guarantees the presence of ultrahigh energy neutrinos.
Over megaparsec distances, UHE nucleons interact with the 2.7K 
microwave background radiation to produce the delta resonance,
which decays to a pion and a nucleon. Charged pions decay to 
neutrinos and muons, followed by muon decay, leading to fluxes of
electron and muon neutrinos. Beginning with the pioneering work of Ref. 1, 
and continuing to recent improvements\cite{hs,ess}, there are quantitative
predictions for fluxes of neutrinos with $E_\nu\sim 10^8$ GeV and
higher energies.

At somewhat lower energies, one expects that sources of TeV photons
such as active galactic nuclei
should also produce neutrinos. The attenuation length of the
photon effectively cuts off the photon energy at $E_\gamma\sim
1$ TeV for sources farther than
100 Mpc from the Earth. The long interaction length
of neutrinos means that neutrino attenuation from distant sources
is not an issue over astronomical distances due to low densities.

A common feature to neutrino flux models is that roughly two
muon neutrinos/antineutrinos appear for each electron neutrino/antineutrino
at the production site. (In the following discussion, we do not distinguish
between $\nu$ and $\bar{\nu}$.) In the past few years, the SuperKamiokande
experiment\cite{superk} together with the SNO\cite{sno} 
results point to physics beyond the
standard model:
neutrino masses
and mixing. Parameters in the mixing matrix lead fairly robustly to an
equal distribution of fluxes of neutrinos between the three neutrino 
flavors\cite{ahluwalia}:
\begin{equation}
F_{\nu_e}=F_{\nu_\mu}=F_{\nu_\tau}
\end{equation}
after propagating over astronomically large distances. UHE
tau neutrino interactions with nucleons in the Earth and atmosphere
may provide signals of this phenomenon. Tests for deviations from
the standard model ultrahigh energy cross section are
another probe of non-standard model physics with UHE neutrinos.

The neutrino energy regime 
determines the detection method. For $E_\nu\leq 10^6$
GeV, underground detectors can be used to record upward muons and electro%
magnetic/hadronic showers. Experiments include water or
ice Cherenkov detectors like AMANDA\cite{amanda},
NESTOR\cite{nestor}, ANTARES\cite{antares} and
IceCube\cite{icecube}.
Attenuation of the neutrino flux in Earth 
is important, especially the difference between $\nu_\mu$ and $\nu_\tau$
attenuation\cite{halzen,us,bottai}.
In the energy range above $10^8-10^9$ GeV, detection of
horizontal air showers and radio detection are used to determine
the UHE $\nu$ fluxes and to probe the neutrino cross 
section\cite{rice,auger,owl,euso,tos,afgs}. 
At intermediate
energies, the double-bang signal of both production and decay vertices
of the $\nu_\tau\rightarrow \tau\rightarrow \nu_\tau X$ process
can be used to identify $\nu_\tau$s\cite{learned}.
Other detection possibilities include using the Earth as a tau
neutrino converter, then searching for the shower from the decay of the
emerging tau\cite{fargion}.

We present results in two of these examples: upward neutrino fluxes
and their detection, and the potential for limits on the non-standard
neutrino cross section from the OWL\cite{owl}
proposed detector.
In this latter case, the non-standard model considered is mini-black
hole production and decay\cite{bh,emr,shrock,bh2,ringwald,muniz,ralston},
a feature of models with TeV-scale compactification
of extra dimensions\cite{dimop}. 
Before going into detail of these two examples,
we review the theoretical status of the neutrino cross section at
high energies.

\section{UHE Neutrino Cross Section}

The neutrino-nucleon 
cross section is straightforward to calculate in the standard
model\cite{quiggtext}. The charged current cross section is 
\begin{equation}
{d^2\sigma\over dx\ dQ^2}={G_F^2\over \pi}\Biggl({M_W^2\over Q^2+M_W^2}
\Biggr)^2 \cdot\Bigl[q(x,Q)+(1-y)^2\bar{q}(x,Q)\Bigr]
\end{equation}
for $W$ boson momentum squared $q^2= -Q^2$, lepton energy
difference $q^0=E-E^\prime=\nu$, Bjorken $x=Q^2/(2M\nu)$ and $y=\nu/E$
with nucleon mass $M$ and $W$ mass $M_W$.
The parton distribution function combinations $q(x,Q)$ and $\bar{q}(x,Q)$
are detailed in, for example, Ref. 31.
The cross section depends crucially on the parton distribution 
functions\cite{andreev}.
As $\ln(Q^2)$ gets large, the parton distributions at small-$x$ increase, 
however,
the $W$ boson propagator decreases with increasing $Q^2$. At sufficiently
high energies, the value of $Q^2$ saturates at $Q^2\sim M_W^2$, so the
relevant values  of $x$ for a given incident neutrino energy $E$ is
approximately
\begin{equation}
x\sim {M_W^2\over 2 M E}\ .
\end{equation}
This translates to $x\sim 10^{-2}$ for $E\sim 10^6$ GeV, and
$x\sim 10^{-8}$ for $E\sim 10^{12}$ GeV.

Measurements of the parton distribution functions have been made down
to $x\sim 10^{-6}$ at $Q^2\sim 0.1$ GeV$^2$ at HERA\cite{hera}.
At a more relevant $Q^2\sim M_W^2$, parton distributions have been measured
to $x\sim 10^{-3}$\cite{dzero}.

Beyond the measured regime, one is guided by theoretical arguments concerning
the small-$x$ parton distribution functions and their QCD evolution.
As discussed in detail in Ref. 35, for $xg(x,Q_0)\sim x^{-\lambda}$
at $x\ll 1$, then $xg(x,Q)\sim x^{-\lambda}$ for $\lambda\gsim 0.3$.
At small values of $x$, the quark distribution functions are
dominated by the quark sea, which come from $g\rightarrow q\bar{q}$, so
the sea distributions have the same 
power law dependence: $xq(x,Q)\sim x^{-\lambda}$.

Modern parton distribution function fits have $\lambda\sim 0.3-0.4$,
and extrapolations to small-$x$ by this power law agree well with
perturbative QCD evolution, 
as demonstrated, for example, by Gl\"uck, Kretzer and
Reya in Ref. 36. Kwiecinski, Martin and Stasto have used a
BFKL-type evolution at small-$x$ and obtain cross sections that are
in agreement with power law extrapolations\cite{kms}.

At very small values of $x$, recombination processes such as $gg\rightarrow
g$ should become important, ultimately
leading to a flattening of the neutrino cross
section as a function of energy. The recombination regime is unimportant
for neutrino energies up to $10^{12}$ GeV, though it should become important
at even higher energies\cite{sterman}.

\begin{figure}
\centerline{\psfig{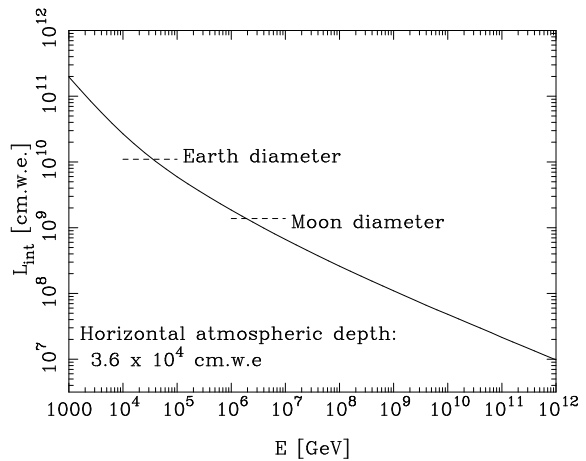}}
\vspace*{8pt}
\caption{The standard model neutrino interaction length in 
cm.w.e.\,=\,g/cm$^2$,
with reference distances through the Earth and the Moon.}
\end{figure}

Fig. 1 shows the neutrino interaction length, proportional to the
inverse of the cross section:
\begin{equation}
{\cal L}_{int}={1\over N_A \sigma_{\nu N}}
\end{equation}
with $N_A$ equal to Avogadro's number. For neutrino energies above
a few tens of TeV, the Earth's diameter is larger than the attenuation
length. Even for neutrinos with $E=10^{12}$ GeV, the horizontal atmospheric
depth is short compared to the neutrino interaction length.

\section{Upward Neutrinos in Underground Detectors}

In our discussion of upward neutrinos, we restrict our attention to
neutrino energies below 10$^6$ GeV\cite{us}. 
For the upward neutrino flux, depending
on the nadir angle, attenuation due to neutrino passage through the
Earth is important. Neutrino oscillations of $\nu_\mu\leftrightarrow
\nu_\tau$ lead to  the arrival of equal fluxes of $\nu_\mu$ and $\nu_\tau$
at the surface of the Earth, however, it has been noted that the
attenuation of the two species is quite different due to the short
lifetime of the $\tau$\cite{halzen,us,bottai}.

Muon and tau neutrinos have the same neutral current and charged current
cross sections above 1 TeV. (At 1 TeV, the charged current cross section
for $\nu_\tau\rightarrow \tau$ is about 5\%  lower than the 
$\nu_\mu$ charged current (CC) cross section due to the $\tau$ mass. At
100 GeV, the $\nu_\tau$ CC cross section is about 25\% smaller\cite{kretzer}.)
The difference in the attenuation of $\nu_\mu$ and $\nu_\tau$ fluxes
has to do with the fact that muons lose energy as they traverse a
medium before they decay. Electromagnetic energy loss for muons is significant,
so the neutrinos from muon decay have much  lower energies than
the parent muon energy. On the other hand, electromagnetic energy loss
for $\tau$'s is important only for tau energies above $\sim 10^8$ 
GeV\cite{drss}. The muon neutrino and tau neutrino transport equations are
\begin{eqnarray}
&&{\partial F_{\nu_{\mu}}(E,X)\over \partial X}=
-{F_{\nu_{\mu}}(E,X)\over {\cal L}
_\nu^{int}(E)} \\ \nonumber
&& \quad\quad
+\int_E^\infty dE_y \,G^{\nu_\mu\rightarrow \nu_\mu}(E,E_y,X)
\\ \nonumber
&&{\partial F_{\nu_{\tau}}(E,X)\over \partial X}=
-{F_{\nu_{\tau}}(E,X)\over {\cal L}
_\nu^{int}(E)} \\ \nonumber &&
+\int_E^\infty dE_y \,[G^{\nu_\tau\rightarrow \nu_\tau}(E,E_y,X)
%\\ \nonumber
%&&\quad\quad\quad\quad\quad\quad\quad  
+G^{\tau\rightarrow \nu_\tau}(E,E_y,X)\,]
\\ \nonumber
&&{\partial F_{{\tau}}(E,X)\over \partial X}\simeq
%-{F_{{\tau}}(E,X)\over {\cal L}_\tau^{int}}
-{F_{{\tau}}(E,X)\over {\cal L}_\tau^{dec}}
\\ \nonumber && \quad\quad
+\int_E^\infty dE_y\,%[ G^{\tau\rightarrow \tau}(E,E_y,X)
%+ 
G^{\nu_\tau\rightarrow \tau}(E,E_y,X)%\,]\ ,
\end{eqnarray}
where, for example,
$$
 G^{\nu_\tau\rightarrow \nu_\tau}(E,E_y,X)
=\Biggl[ {F_{\nu_\tau}(E_y,X)\over {\cal L}_\nu^{int}}
\Biggr]{dn^{NC} \over dE}(E_y,E)\ .
$$
in terms of the neutrino flux $F_{\nu_\tau}$ and the cross section
normalized neutral current energy distribution $dn^{NC}/dE$. We have made
the approximation that tau energy loss is negligible, and that
only the decay of the tau is relevant, neglecting tau CC interactions.

\begin{figure}
\centerline{\psfig{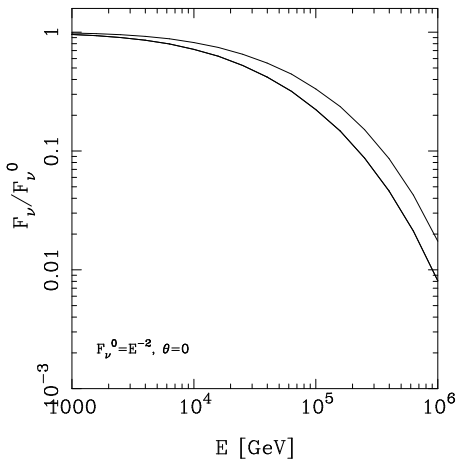}}
\vspace*{8pt}
\caption{The ratio of the attenuated flux to incident flux for
$\nu_\tau$ (upper curve) and $\nu_\mu$ (lower curve) for $F_\nu^0\sim
E^{-2}$ at nadir angle 0.}
\end{figure}

\begin{figure}
\centerline{\psfig{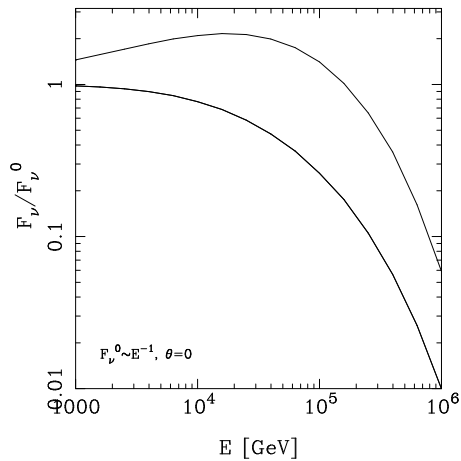}}
\vspace*{8pt}
\caption{The ratio of the attenuated flux to incident flux for
$\nu_\tau$ (upper curve) and $\nu_\mu$ (lower curve) for $F_\nu^0\sim
E^{-1}$ at nadir angle 0.}
\end{figure}

The feed-down of neutrinos from $\nu_\tau\rightarrow \tau$ is evident
in comparison of the attenuated $\nu_\tau$ flux with the $\nu_\mu$ 
flux\cite{us,bottai}.
In Figs. 2 and 3 we show the ratio of the attenuated flux to the incident
flux for $\nu_\tau$ (upper curve) and $\nu_\mu$ (lower curve) at nadir
angle 0 for two different fluxes, parameterized by
\begin{equation}
F_\nu^0=E^{-\gamma-1}/(1+E/E_0)^\alpha
\end{equation}
for $E_0=10^8$ GeV.
In Fig. 2, $\gamma=1$ and $\alpha=0$, and in Fig. 3, $\gamma=0$ and $\alpha=2$.
The tau neutrino pile-up is more pronounced for the flatter fluxes.
At larger nadir angles, the pileup reduces because the number of interactions
reduces, but the attenuation is also less.

Experimentally, one would like to exploit the effect of the tau neutrino
pile-up and to identify the tau neutrino component of the flux.
In a manner similar to the lower energy comparison of the neutral current
to charged current event rates\cite{sno}, for neutrinos with energies
above 1 TeV, one can compare the upward electromagnetic/hadronic shower event
rate from
\begin{eqnarray}
&& \nu_\tau N\rightarrow \tau+{\rm hadrons},\ \ \tau\rightarrow \nu_\tau+
{\rm hadrons}\\ \nonumber
&& \nu_\tau N\rightarrow \tau+{\rm hadrons},\ \ \tau\rightarrow \nu_\tau+
e+\nu_e\\ \nonumber
&& \nu_{\tau,\mu,e} N\rightarrow \nu_{\tau,\mu,e}+{\rm hadrons}\\ \nonumber
&& \nu_eN\rightarrow e+{\rm hadrons}
\end{eqnarray}
with the upward muon event rate from
\begin{eqnarray}
&& \nu_\mu N\rightarrow \mu+X \\ \nonumber
&& \nu_\tau N\rightarrow \tau+X,\ \ \tau\rightarrow \nu_\tau+\mu+\nu_\mu \ .
\end{eqnarray}
We have made a detailed comparison in Ref. 12. By measuring the upward
shower and muon event rates with an energy threshold of 10 TeV, one should
be able to see the effect of the tau neutrino component of the incident
neutrino flux. This is relatively independent of the
{\it a priori} unknown spectrum of the incident flux, except for
the overall normalization (so that there are enough events).

\begin{figure}
\centerline{\psfig{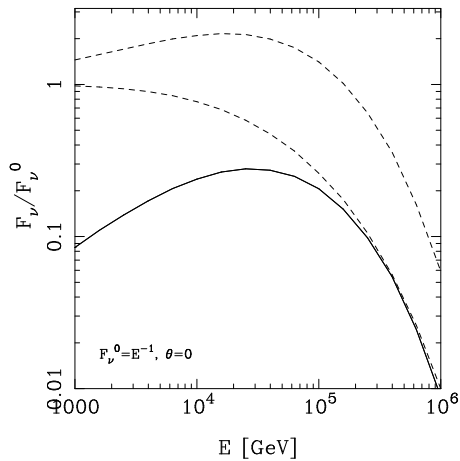}}
\vspace*{8pt}
\caption{The ratio of the attenuated flux to incident flux for
$\nu_\tau$ (upper dashed curve) and $\nu_\mu$ (lower dashed curve) and
secondary $\nu_\mu$ (solid curve)
for $F_\nu^0\sim
E^{-1}$ at nadir angle 0.}
\end{figure}

\begin{figure}
\centerline{\psfig{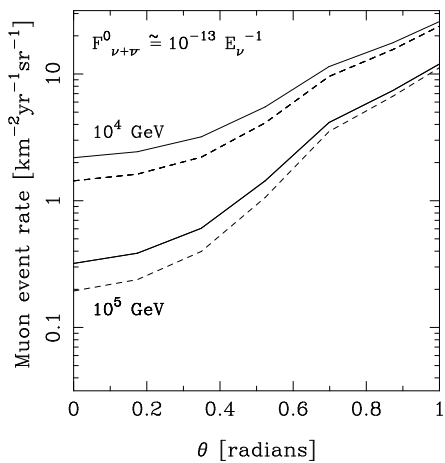}}
\vspace*{8pt}
\caption{Upward muon event rate for $F_\nu^0\sim 10^{-13}$ (GeV/$E$)
(cm$^2$s\,sr\,GeV)$^{-1}$ without secondary neutrinos (dashed curve) and
with secondary neutrinos (solid curve) for $E_\mu\geq 10^4,\ 10^5$ GeV.}
\end{figure}

An interesting possibility for a class of input neutrino fluxes is the
enhancement of the upward muon rate from $\nu_\tau\rightarrow\tau\rightarrow
\nu_\mu$ where the final muon neutrino interacts to produce a 
muon\cite{beacom}.
The muon neutrinos from tau decays are called ``secondary neutrinos."
There is incremental energy loss at each of the processes: CC production
of the tau, then muonic decay of the tau. This results in a contribution
to the muon neutrino flux\cite{secondaries}, 
as seen in Fig. 4 for $F\sim 1/E$ for
nadir angle 0. The upper dashed line is $F_\nu/F_\nu^0$ for the $\nu_\tau$ 
flux, the lower dashed line is for $\nu_\mu$ excluding the secondary
neutrinos, while the solid line is the secondary muon neutrino flux.
The implications for the upward muon rate are shown in Fig. 5.
The secondary neutrino flux for the $1/E^2$ incident flux is relatively
less important. For less steep fluxes than $1/E$, one expects that the
secondary flux will be enhanced. 

\section{Horizontal Air Shower Signals of Mini-black Holes}

As Fig. 1 indicates, the standard model neutrino interaction length
is large compared to the horizontal atmospheric depth (zenith angle
$90^\circ$ at sea level). Unlike strongly interacting particles,
neutrinos can penetrate deep into the atmosphere. Horizontal air
showers initiated deep in the atmosphere are a signal of neutrinos.
Non-observation of an enhancement of horizontal air shower event rates
limit a combination of the neutrino-nucleon cross section and the 
neutrino flux\cite{tos,afgs}. Using the cosmogenic neutrino flux from
cosmic ray scattering with the 2.7K photons, it is possible to set
limits on non-standard contributions to the cross section.

\begin{figure}
\centerline{\psfig{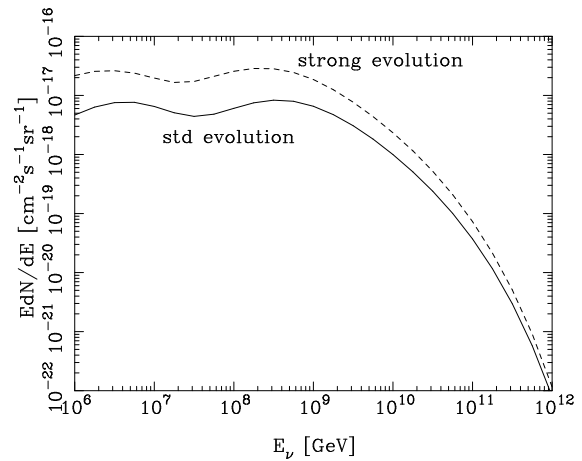}}
\vspace*{8pt}
\caption{The cosmogenic electron neutrino plus antineutrino flux
of Engel, Seckel and Stanev.$^3$}
\end{figure}

The cosmogenic neutrino flux, for the sum of electron neutrino plus
antineutrino, evaluated by Engel, Seckel and Stanev\cite{ess}
(ESS) is shown in Fig. 6. Their results are shown for two models
of source evolution: one parameterized like $(1+z)^3$ for redshift $z<1.9$
(standard evolution),
and the other scaling like $(1+z)^4$ for the same redshift (strong evolution).

Discussed here are contributions from mini-black holes, predicted by models
with the Planck scale in the TeV range and a number of extra 
dimensions\cite{dimop},
and the potential of an orbiting observatory to detect the enhanced
rate of horizontal air showers\cite{bhus}. There is an extensive
literature on black hole production and detection by
neutrino telescopes\cite{afgs,bh,emr,shrock,bh2,ringwald,muniz,ralston}.
It appears that enhancements
of the cross section from $t$-channel Kaluza-Klein graviton exchange
is not an important effect\cite{emr,shrock}.

Black hole production in these models is determined by the geometrical
cross section, depending on the Schwarzschild radius. The Schwarzschild
radius $r_S$, assuming $n$ compactified dimensions and Planck scale
$M_D$, is given by\cite{colliderth}
\begin{equation}
r_S =\frac{1}{M_D}
\left[
\frac{M_{\rm BH}}{M_D}
\left(
\frac{2^n \pi^{\frac{n -3}{2}}\,\Gamma\left( \frac{3+n}{2}\right)}
{2+ n }
\right)
\right]^{\frac{1}{1+ n}}\ .
\end{equation}
Here, $M_{\rm BH}$ is the mass of the black hole. The geometrical
cross section for neutrino interaction with parton $j$ is then 
\begin{equation}
\hat\sigma (\nu j\rightarrow BH) = \pi r_S^2(M_{BH}=\sqrt{\hat s}) 
\theta (\sqrt {\hat s} - M_{BH}^{\rm min})
\end{equation}
for $\hat{s}$ equal to the neutrino-parton center of mass energy.
Eq. (10) enforces the requirement that $\hat{s}$ be larger than some minimum
energy-squared, certainly larger than the Planck scale. Convoluting
with the parton distribution functions $f_i$ after setting $\hat{s}=x\, s$
give a non-standard model contribution to the neutrino-nucleon cross section:
\begin{equation}
\sigma (\nu N \to {\rm BH}) = 
\sum_i \int_{\frac{(M_{BH}^{\rm min})^2}{s}}^1 dx \,\, 
\hat\sigma_i^{BH}(xs)\,\, f_i(x,Q^2)\ .
\end{equation}
The black hole is expected to rapidly evaporate into standard model
particles, and we
assume that all of the energy of the black hole goes into the shower it
produces\cite{colliderth}.

Experimental constraints on deviations from Newtonian gravity
exclude $n=1$ and $M_D\sim 1$ TeV\cite{afgs}. For $n=2$, $M_D>3.5$ TeV
is consistent with tests of the inverse square law,\cite{neq2} while
for larger $n$, there are no constraints from tests of Newtonian
gravity for $M_D\sim 1$ TeV. Collider experiments put a lower limit
on $M_D$ at approximately 1 TeV\cite{afgs,collider}. 

\begin{figure}
\centerline{\psfig{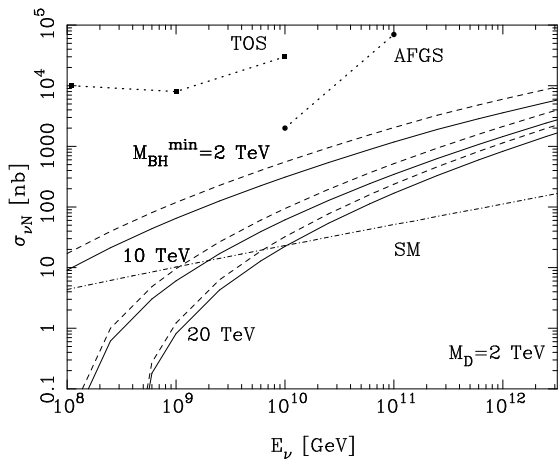}}
\vspace*{8pt}
\caption{The black hole production cross section for $n=4$ (solid) and
$n=6$ (dashed) for several values of the minimum black hole mass, for
a Planck scale of $M_D=2$ TeV. The standard model cross section is
also shown (SM) with the dot-dashed line. The limits of Tyler, Olinto
and Sigl$^{18}$ (TOS)  and Anchordoqui {\it et al.}$^{19}$
(AFGS) are shown with the
dotted lines.}
\end{figure}

Fig. 7 shows a representative set of parameter choices for the
number of extra dimensions $n$ and the minimum black hole mass, given
a Planck scale of $M_D=2$ TeV. Also shown are curves representing
upper limits on the cross section from Tyler, Olinto and Sigl\cite{tos}
(TOS) and Anchordoqui, Feng, Goldberg and Shapere\cite{afgs} (AFGS),
were they have used cosmogenic fluxes similar to that of ESS. These upper 
limits
come from the AGASA and Fly's Eye limits on an excess of horizontal
air shower events.

The largest enhancement of the cross section comes at the highest energies.
Detectors optimized to this energy range will be able to put the
most stringent limits on the cross section. The proposed Orbiting
Wide-Angle Light-Collectors Experiment\cite{owl} (OWL) and the Extreme Universe
Space Observatory\cite{euso} (EUSO) are two such detectors. The OWL
experiment would involve two satellites with photodetectors, orbiting
640 km above the Earth. The EUSO experiment would be located on the
International Space Station, 380 km above the Earth. The conversion factor
from the 2 telescopes to one at a reduced altitude amount to a reduction
of the event rate by a factor of about 0.2 compared to OWL.

\begin{figure}
\centerline{\psfig{file=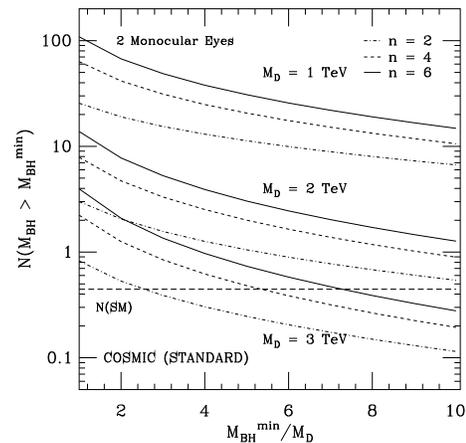,width=6cm}}
\vspace*{8pt}
\caption{The event rate per year for black hole production for
2 monocular eyes for OWL from the ESS cosmogenic electron neutrino flux
(standard evolution), for $n=2,4,6$ extra dimensions, $M_D=1,2,3$ TeV
as a function of $M_{\rm BH}^{\rm min}/M_D$. Also shown is the standard
model event rate.}
\end{figure}

The threshold for OWL detection of horizontal air showers is on the order
of $10^{10}$ GeV. By rescaling the effective aperture for OWL detection
of electron neutrinos by the ratio of the standard model plus black hole
cross section to the standard model neutrino-nucleon cross section, one
can arrive at event rate predictions given the ESS incident electron
neutrino flux. The range of electron neutrino induced
event rates goes from 100 events per year
for $n=6$ and $M_D=M_{\rm BH}^{\rm min}=1$ TeV to less than one event
per year for $M_D=3$ TeV and $M_{\rm BH}^{\rm min}=10\cdot M_D$ for the
standard evolution of the cosmogenic flux
as shown in Fig. 8. The standard model event rate
is about 0.5 events/year. The strong evolution of the cosmogenic
flux gives a factor of 2 enhancement for both signal and background.
The OWL rates are about a factor of 20 times larger than the IceCube
downward contained event rates for an energy threshold of $10^8$ GeV. 
IceCube will optimally detect fluxes with large low energy components,
{\it e.g.}, $F_\nu\sim 1/E^2$\cite{muniz}. 

If all three neutrino flavors are included in the event rate calculation,
the standard model rates increase by a factor of about 2, while the
black hole rates are increased by a factor of 3 since the black hole
cross section is flavor-blind. (The standard model rate depends on how
much of the incident neutrino energy is translated to the shower in the
final state.) Including these factors, one finds that OWL would be able
to probe the fundamental Planck scale up to 3 TeV for $n\geq 4$ even
with one year of data taking.

\section{Final Remarks}

Described above are two ways to search for nonstandard
model physics with ultrahigh energy neutrinos: searching for evidence
of $\nu_\mu\rightarrow \nu_\tau$ oscillations with tau appearance
in upward muon and upward shower rates,
and mini-black hole production in neutrino induced horizontal air showers.
Nonstandard model physics may be manifest in the flux of neutrinos, through
exotic particle decay to neutrinos\cite{sigl}. A wide range of experiments:
underground Cherenkov detectors, surface arrays and satellite 
light collectors cover a huge range of energies from GeV to 10$^{12}$ 
GeV incident neutrinos. The 
view from the ultrahigh energy neutrino window on 
astrophysics and particle physics should provide interesting tests of
the standard model and its extensions.

\section*{Acknowledgments}

The work of S.I.D. has been supported in part by National Science
Foundation Grant No. 0070998. The work of I.S. has been
supported in part by the DOE under contracts No. DE-FG02-95ER40906
and DE-FG02-93ER40792. The work of M.H.R. has been supported
in part by the DOE under contract No. FG02-91ER40664.

\end{document}